\documentclass[journal,twoside,web]{IEEEtran}  

\usepackage{cite}\usepackage{hyperref}
\usepackage{amsmath,amssymb,amsfonts}
\usepackage{algorithmic}
\usepackage{graphicx}
\usepackage{textcomp}
\usepackage{amsmath,amssymb,bm,bbm,mathrsfs,amscd}
\usepackage{calc}
\usepackage{color}
\usepackage{dsfont}
\usepackage{graphicx}
\usepackage{epstopdf}
\usepackage{epsfig}
\usepackage{tikz}
\usepackage{amsmath}
\usepackage{amsfonts}
\usepackage{amssymb}
\usepackage{rotating}
\usepackage{mathtools}
\def\qed{\hfill \vrule height 7pt width 7pt depth 0pt\medskip}
\usepackage{color}
\usepackage{subfig}
\usepackage{enumerate,pgfplots}
\newtheorem{theorem}{Theorem}
\newtheorem{definition}{Definition}
\newtheorem{proposition}{Proposition}
\newtheorem{lemma}{Lemma}
\newtheorem{corollary}{Corollary}
\newtheorem{example}{Example}
\newtheorem{remark}{Remark}

\newcommand{\ba}{\begin{array}}
	\newcommand{\ea}{\end{array}}

\newcommand{\mb}{\boldsymbol}
\newcommand{\be}{\begin{equation}}
\newcommand{\ee}{\end{equation}}

\newcommand{\ds}{\displaystyle}

\newcommand{\eps}{\varepsilon}

\newcommand{\mc}{\mathcal}

\newcommand{\ov}{\overline}

\def\1{\boldsymbol{1}}

\newcommand{\E}{\mathbb{E}}

\newcommand{\R}{\mathbb{R}}

\newcommand{\F}{\mathbb{F}}

\newcommand{\pmut}{1\!-\!\rho}
\newcommand{\pint}{\rho}
 \newcommand{\ldrift}{\ov D}

\renewcommand{\l}{\left}\renewcommand{\r}{\right}

\newcommand{\xb}{{\bf x}}

\renewcommand{\P}{\mathbb{P}}

\renewcommand{\Re}{\R\mathrm{e}}

\DeclareMathOperator*{\argmax}{argmax}

\def\E{\mathbb{E}}
\def\R{\mathbb{R}}

\def\P{\mathbb{P}}

\tikzstyle{v_c}=[circle, draw,inner sep=2pt, minimum width=12pt, color=blue]
\tikzstyle{v_a}=[circle, draw,inner sep=2pt, minimum width=12pt, color=red]
\tikzstyle{edge} = [draw,thick,-,font=\small ]
\tikzstyle{label} = [draw,fill=black,font=\normalsize]

\def\F{{\mathcal F}}

\def\BibTeX{{\rm B\kern-.05em{\sc i\kern-.025em b}\kern-.08em
		T\kern-.1667em\lower.7ex\hbox{E}\kern-.125emX}}
\title{\LARGE \bf 
An invariance principle based concentration result for large-scale stochastic pairwise interaction network systems}
\author{
	Giacomo~Como,~\IEEEmembership{Member,~IEEE,}
	Fabio Fagnani,~\IEEEmembership{Member,~IEEE,}
	and Sandro Zampieri~\IEEEmembership{Fellow,~IEEE}  
	\thanks{G.~Como and F.~Fagnani are with the  Department of Mathematical Sciences ``G.L.~Lagrange,'' Politecnico di Torino, 10129 Torino, Italy  (e-mail: {\{giacomo.como;\,fabio.fagnani\}@polito.it}). G.~Como is also with the Department of Automatic Control, Lund University, 22100 Lund, Sweden. S.~Zampieri is with the Department of Information Engineering, University of Padua, via Gradenigo 14/b, Padova, Italy (e-mail: zampi@dei.unipd.it). }
	\thanks{This work was partially supported by a MIUR  Research Project PRIN 2017 ``Advanced Network Control of Future Smart Grids'' (http://vectors.dieti.unina.it) and by the Compagnia di San Paolo.}%
}

\begin{document}

	\maketitle
	\thispagestyle{empty}
	\pagestyle{empty}
	
\begin{abstract}

We study stochastic pairwise interaction network systems whereby a finite population of agents, identified with the nodes of a graph, update their states in response to both individual mutations and pairwise interactions with their neighbors. The considered class of systems include the main epidemic models ---such as the SIS, SIR, and SIRS models---, certain social dynamics models ---such as the voter and anti-voter models---, as well as evolutionary dynamics on graphs. 
 Since these stochastic systems fall into the class of finite-state Markov chains, they always admit stationary distributions.
We analyze the asymptotic behavior of these stationary distributions in the limit as the population size grows large while the interaction network maintains certain mixing properties. 
Our approach relies on the use of Lyapunov-type functions to obtain concentration results on these stationary distributions.
Notably, our results are not limited to fully mixed population models, as they do apply to a much broader spectrum of interaction network structures, including, e.g., Erd\"os-R\'enyi random graphs. 
\end{abstract}


\section{Introduction}\label{sec:introduction}
Pairwise Interaction Network models (PIN models) constitute a class of network systems whereby a finite ---but possibly very large--- population of agents update their states asynchronously, according to stochastic rules that account for both individual mutations and pairwise interactions between neighbor agents \cite{Donnelly.Welsh:83,Liggettbook,Liggettbook2}. 
The class of PIN models encompasses the main microscopic epidemic models ---such as SIS, SIR, or SIRS models over networks \cite{Nowzari2016}---, certain social dynamics models ---such as the voter, anti-voter, and the Axelrod models \cite{Cox:89,Donnelly.Welsh:84,Castellano.ea:09}--- as well as evolutionary dynamics on graphs \cite{Lieberman.ea:05,Nowak:06}. 

Despite the significant attention that PIN models have received over the years, their theoretical understanding on general interaction patterns still presents considerable gaps. With some notable exceptions ---including the analysis of the extinction time in network SIS epidemic models \cite{Ganesh.ea:04} or of size of the population that eventually becomes infected in network SIR epidemic models--- the majority of the literature on the analysis of epidemic models has concentrated on mean-field models, studying the epidemic as a system of ordinary differential equations (ODEs)\cite{Kermack.McKendrick:1927,Lajmanovich.Yorke:1976,Mei.ea:2017,Alutto.ea:24}. 
In fact, under proper technical assumptions, the Kurtz theorem \cite{Kurtz:70} ensures that the solutions of such mean-field ODEs {well approximate the evolution of PIN models on complete interaction graphs of increasing size over bounded time horizons,} 
see, e.g., \cite[Chapter 10]{Sandholm:10} or \cite[Chapter 5]{DM:2010}. Recent works explore extensions of these asymptotic results to sequences of both sparse and dense graphs \cite{Janson.ea:2014,Rossi.ea:19,KHT:2022,Ravazzi.ea:23} and to continuous state space \cite{Como.Fagnani:11}. 

While the Kurtz theorem provides a deterministic approximation of the transient dynamic behavior of fully mixed stochastic network systems in the large-scale limit, the infinite-horizon behavior of such systems is best captured by characterizing their stationary distributions for finite population sizes and possibly analyzing their convergence times to stationarity \cite[Chapter 11]{Sandholm:10}. When the considered stochastic network system lacks ergodicity due, e.g., to the existence of absorbing configurations (such as the all-susceptible configuration in the aforementioned SIS and SIRS epidemic models), a common approach consists in introducing a (small) noise term that makes the system ergodic and characterizing the dependence of the stationary distribution on the noise level \cite{Kandori.ea:93,Ellison:93,Fudenberg.Levine:98}. {Explicit analytical expressions are rare and essentially limited to reversible systems \cite{Blume:93,Blume:97}. However, it is often possible to study the asymptotic behavior of such stationary distributions in the limit of large population (and vanishing noise) \cite[Chapter 12]{Sandholm:10}.} 
By studying this (double) limit, one can often show that the probability mass of every sequence of invariant distributions of these network systems concentrates on either a single or a small subset of population states, thus leading to equilibrium selection of so-called stochastically stable population states \cite{Young:93,Benaim.Weibull:03}. Notably, {for mean-field models,} the set of such stochastically stable population states have been proved to be as a subset of the so-called Birkhoff center of the corresponding mean-field ODE, defined as the closure of the set of its recurring points: c.f.~\cite{Benaim:98,Benaim.LeBoudec:11} and \cite[Theorem 12.6.2]{Sandholm:10}. While such Birkhoff center includes equilibrium points, cycles, and more complex limit sets of the mean-field ODE, regardless of their stability, in some special cases this analysis has been refined to rule out unstable equilibrium points of the mean-field ODE  \cite[Theorem 12.6.4]{Sandholm:10}.

In this paper, we focus on PIN models over general interaction networks and show that some of the results obtained for mean-field PIN models can actually be extended {much beyond complete interaction graphs.}
This is achieved by employing Lyapunov function arguments tailored to the PIN models, enabling us to establish a concentration result for their stationary probability distributions.
Specifically, we demonstrate that if the mean-field ODE associated with the PIN model admits a class-$\mc C^2$ global Lyapunov function $V$, then in the large-scale limit every sequence of stationary distributions concentrates on the set of zeros of the time-derivative $\dot V$ of such Lyapunov function. This occurs under the assumption that the interaction network satisfies a specific topological property, referred to as asymptotic total mixing.
As we shall illustrate, in addition to complete graphs, random graphs such as the Erd\"os-R\'enyi model also enjoy this property. As a consequence, our approach applies to a significantly broader spectrum of interaction networks beyond mean-field models.
We then generalize our results to account for cases where a class-$\mc C^2$ Lyapunov function is available only for suitable noisy perturbations of the considered PIN model but not necessarily for the original one, as is the case for the SIRS epidemic model.

%

We wish to point out that, while the use of Lyapunov functions for Markov chains is standard in the context of stochastic stability  (c.f.~the celebrated Forster theorem \cite{Foster:53,MeynTweede:93} for positive recurrence of irreducible Markov chains with infinite state space), the technical novelty of our approach resides in  using Lyapunov functions to prove concentration results for PIN models. Indeed, the latter are finite-state Markov chains, so that existence of stationary distributions is always guaranteed and their uniqueness is implied by irreducibility \cite{Norris:98}, whereas the interest is rather on concentration of their stationary distributions in the large-scale limit. In this sense, rather than stability analysis, our results are to be considered as a sort of invariance principle for large-scale stochastic PIN models. 

The rest of the paper is organized as follows. The last paragraph of this Introduction gathers some notational conventions. In Section \ref{sec:PIN-model}, PIN models are introduced. In Section \ref{sec:main-results}, we first present the important concepts of limit drift and mean-field Lyapunov function (Section \ref{sec:limitdrif+lyapunov}) as well as of asymptotically totally mixing network (Section \ref{sec:ATM-networks}), and then formally state our main results on concentration properties of PIN models in the large-scale limit (Section \ref{sec:concentration-results}). In Section \ref{sec:examples}, {various examples are discussed,}
including forgetful PIN models (Section \ref{sec:forgetful}), binary PIN models (Section \ref{sec:binary}), as well as the SIRS model that does not fit in either of the two categories (Section \ref{sec:nonbinary}). The main technical contributions are in Section \ref{sec:main-concentration}, where first the mean drift of PIN models is analyzed (Section \ref{sec:mean-drift}), then concentration results are developed for PIN models on finite networks (Section \ref{sec:concentration-finite}), and finally such results are applied in order to prove the large-scale limit concentration results (Section \ref{sec:proofs}). Section \ref{sec:conclusion} gathers some concluding remarks while the Appendices contain the proofs of some technical results, including the proof that the Erd\"os-R\'enyi random graph is asymptotically totally mixing.

{\bf Notation} For a finite set $\mc A$, its cardinality is denoted by $|\mc A|$. 
For two finite sets $\mc A$ and $\mc B$, and another, not necessarily finite, set $\F$, we denote by 
$\F^{\mc A}$ and $\F^{\mc A\times\mc B}$ the spaces of vectors $z$ of dimension $|\mc A|$ and of matrices $Z$ of dimension $|\mc A|\times|\mc B|$, with entries $z_i$ in $\F$ indexed by the elements $i$ of $\mc A$ and $Z_{ij}$ in $\mc F$ indexed by the pairs $(i,j)$ in $\mc A\times\mc B$, respectively. 
{ For a vector $a$ in $\R^{\mc A}$ the symbols $\|a\|_\infty$, $\|a\|_2$, and $\|a\|_1$ stand for the standard $l_\infty$, $l_2$, and $l_1$ norms. 
For a matrix $A$ symbol $\|A\|_\infty$ means the $\infty$-norm of its vectorization, namely the maximum absolute value of its entries. }
The identity matrix is denoted by $I$, the all-one vector by $\1$, while $\delta^j$ stands for the vector with all entries equal to zero except for the $j$-th that is equal to $1$. 
For a vector $z$ in $\mc F^{\mc A}$ and some $i$ in $\mc A$, $z_{-i}$ denotes the vector in $\mc F^{\mc A\setminus\{i\}}$ obtained from $z$ by removing its $i$-th entry. 
The simplex of probability vectors over a finite set $\mc A$ is denoted by 
$$\mc P(\mc A)=\left\{z\in\R_{+}^{\mc A}:\,||z||_1=1\right\}\,$$
where $\R_{+}$ denotes the set on non-negative real numbers.

A (finite, directed) graph $\mc G=(\mc V,\mc E)$ is the pair of nonempty finite set of nodes $\mc V$ and of a set of links $\mc E\subseteq\mc V\times\mc V$. We shall always assume that there are no self-loops, i.e., that $(i,i)\notin\mc E$ for every $i$ in $\mc V$. We shall refer to a graph $\mc G=(\mc V,\mc E)$ as undirected if $(i,j)\in\mc E$ implies that $(j,i)\in\mc E$ and as nonempty if $\mc E\ne\emptyset$.

\section{PIN models and problem description}\label{sec:PIN-model}
We consider stochastic \emph{pairwise interaction network (PIN) models}, whereby a nonempty finite set of agents $\mc V$ are engaged in repeated interactions  on a nonempty graph $\mc G=(\mc V,\mc E)$. The nodes in $\mc V$ represent the agents and the directed links in $\mc E$ represent the allowed pairwise interactions. 
Precisely, the presence of a link $(u,v)$ in $\mc E$ directed from its tail node $u$ in $\mc V$ to its head node $v\ne u$ in $\mc V$ indicates an influence of the state of agent $v$ on that of agent $u$. 
We shall refer to the graph $\mc G=(\mc V,\mc E)$ as the \emph{interaction pattern} and to $n=|\mc V|$ and $m=|\mc E|$ as the \emph{order} and the \emph{size}, respectively, of the PIN model.

Every agent $u$ in $\mc V$ is endowed with a time-varying \emph{state} $X_u(t)$, taking values in a nonempty finite state set $\mc A$, at every discrete time instant $t=0,1,2,\ldots$. We stack all the agents' states in a vector $X(t)=(X_u(t))_{u\in\mc V}$, to be referred to as the system \emph{configuration}, taking values in the configuration space $\mc X=\mc A^{\mc V}$. 
The PIN model then generates a discrete-time Markov chain $X(t)$ on the configuration space $\mc X$, whereby, for every time instant $t=0,1,2,\ldots$, the next configuration $X(t+1)$ differs from the current one $X(t)$ in at most one entry, corresponding to an agent, which modifies her action as a result of either an  individual mutation or of a pairwise interaction with a neighbor agent in the interaction graph, as described below.  

More precisely, at every time instant $t=0,1,2,\ldots$, we have that:  
\begin{itemize}
\item a pairwise interaction takes place with probability $\pint$, while an individual mutation takes place with probability $\pmut$; 
\item given that an individual mutation takes place, an agent $u$ is chosen uniformly at random from $\mc V$ and gets activated: given its  current state  $X_u(t)=i$ in $\mc A$, agent $u$ modifies it to a new state $X_u(t+1)=j$ in $\mc A$ with conditional probability $P_{ij}$, while the rest of the agents keep their states unaltered, so that $X_{-u}(t+1)=X_{-u}(t)$;  
\item  given that a pairwise interaction takes place, a link $(u,v)$ is activated, chosen uniformly at random from the link set  $\mc E$: given that agent $u$ ---corresponding to the tail node of the activated link--- is in state $X_u(t)=i$ in $\mc A$, and agent $v$ ---corresponding to the head node of the activated link--- is in state $X_v(t)=\ell$ in $\mc A$, agent $u$ modifies its state to a new state $X_u(t+1)=j$ in $\mc A$ with conditional probability $\phi_{ij}(\ell)$, while once again the rest of the agents keep their states unaltered, so that $X_{-u}(t+1)=X_{-u}(t)$.  
\end{itemize}
Throughout, it is always implicitly assumed that $$0\le\pint\le1\,,\qquad\sum_{j\in\mc A}P_{ij}=1\,,\qquad\sum_{j\in\mc A}\phi_{ij}(\ell)=1\,,$$ for every $i$ and $\ell$ in $\mc A$. We shall refer to the row-stochastic matrix $P$ in $\R_+^{\mc A\times\mc A}$ as the \emph{mutation transition matrix} and to the map $\phi:\mc A\to\R_+^{\mc A\times\mc A}$ as the \emph{interaction transition tensor}. We shall refer to the quadruple $(\mc A, \pint, P, \phi)$ as the \emph{parameters} of the considered PIN model.

A probability vector $\mu$ in $\mc P(\mc X)$ over the configuration space $\mc X$ is referred to as a \emph{stationary distribution} for a PIN model if, whenever its initial configuration $X(0)$ has probability distribution $\P(X(0)=\xb)=\mu_{\xb}$ for all $\xb$ in $\mc X$,  the model configuration $X(t)$ has probability distribution $\P(X(t)=\xb)=\mu_{\xb}$ for all $\xb$ in $\mc X$ for every time instant $t=0,1,2,\ldots$. Observe that, since every PIN model is a Markov chain on the finite configuration space $\mc X$, it always admits at least one stationary distribution $\mu$ in $\mc P(\mc X)$. 

A PIN model is referred to as \emph{ergodic} if, for every two configurations $\xb$ and ${\bf y}$ in $\mc X$, there exists a sequence of node and link activations and of corresponding spontaneous mutations and pairwise interactions, respectively, all having positive probability, that lead the PIN configuration from $X(0)=\xb$ to $X(t)={\bf y}$ in a finite number of time steps $t\ge0$. It is well known that the stationary distribution $\mu$ of an ergodic PIN model (and in fact of every ergodic finite-state Markov chain \cite{Norris:98}) is unique and assigns positive probability $\mu_\xb>0$ to every configuration $\xb$ in $\mc X$. Notice that ergodicity is ensured, for instance, provided that the mutation probability $\pmut>0$ is positive and the mutation transition matrix $P$ is irreducible. 

\begin{remark} Observe that the stationary distribution of a PIN model may not be unique if the PIN model is not ergodic.  Although our theory is applicable to all such PIN models, the results that will be presented hold particular significance  for ergodic PIN models. As mentioned in the Introduction, certain PIN models of interest that are not themselves ergodic can be analyzed by introducing a perturbation in the mutation kernel that makes them ergodic and then taking the limit as such perturbation vanishes. Such vanishing noise limit approach has been studied, e.g., in the theory of random perturbations of dynamical systems \cite{Freidlin.Wentzell:12} and then applied within the theory of learning in games \cite{Kandori.ea:93,Ellison:93,Fudenberg.Levine:98}. 
\end{remark}
\medskip 


In the rest of the paper we shall study concentration properties of stationary distributions of PIN models. 
Specifically, let  $\theta:\mc X\to\mc P(\mc A)$ be the function mapping a configuration $\xb$ in $\mc X$ into its  \emph{type} $\theta(\xb)$ in $\mc P(\mc A)$ whose entries
$$\theta_i(\xb):=\frac{1}{n}|\{v\in\mc V\,|\,\xb_v=i\}|\,,\qquad i\in\mc A\,,$$ 
represent the empirical frequencies of the different states in the configuration $\xb$. 
We shall then focus on concentration properties of the \emph{type process} defined as follows
\be\label{type-process}\Theta(t):=\theta(X(t))\,,\qquad t=0,1,\ldots\,.\ee
Notice that, for every discrete time instant $t=0,1,2,\ldots$,  the random variable $\Theta(t)$ takes values in the compact simplex $\mc P(\mc A)$ of probability vectors over the state space $\mc A$. 
In particular, we shall focus on the asymptotic behavior of the distribution of $\Theta(t)$ in the large-scale limit, i.e., as the order $n$ of the PIN model grows large while its parameters $(\mc A, \pint, P, \phi)$ remain fixed. 

\begin{remark}\label{remark:complete}
The type $\Theta(t)$ associated with a PIN model with state space $\mc A$ and order $n$ takes values in a finite subset 
$$\mc P_n(\mc A)=\{\theta(\xb):\,\xb\in\mc X\}\,,$$ of the simplex $\mc P(\mc A)$.
Now, observe that, when the interaction pattern of a PIN model is the complete graph, the type process $\Theta(t)=\theta(X(t))$ is itself a Markov chain with state space $\mc P_n(\mc A)$. 
Since the cardinality 
$$|\mc P_n(\mc A)|=\binom{n+|\mc A|-1}{|\mc A|-1}\le(n+1)^{|\mc A|}\,,$$
grows polynomially fast in the order $n$ while the cardinality $|\mc X|=|\mc A|^n$ of the configuration space $\mc X$ grows exponentially fast, the special case of complete interaction pattern provides a formidable reduction of complexity. In contrast, when the interaction network is not a complete graph, the type process $\Theta(t)=\theta(X(t))$ is in general not Markovian, while the process $X(t)$ always is. This creates a specific technical challenge that will be addressed in the rest of the paper. 
\end{remark}\medskip 


\section{Main results}\label{sec:main-results}
In this section, we present our two main results regarding concentration of stationary distributions of stochastic PIN models in the large-scale limit. 
{We start by introducing a few key concepts related to PIN models. }

\subsection{Limit drift and mean-field Lyapunov functions}\label{sec:limitdrif+lyapunov}
We start by defining the notions of  \emph{limit drift} and \emph{mean-field Lyapunov function} for a PIN model. 

\begin{definition}\label{def:llimdrift}
The \emph{limit drift}  of a PIN model with parameters $(\mc A,\pint, P,\phi)$ is the function $$\ldrift:{\mc P(\mc A)}\to \R^{\mc A}\,,$$ mapping a probability vector $\theta$ into the vector
\be
\label{drift-mf}\ldrift(\theta):= 
\left((\pmut) P^\top+\pint\sum_{\ell\in\mc A}\theta_\ell\phi(\ell)^\top-I\right)\theta\,.\ee
\end{definition}

\begin{definition}\label{def:meanfieldLyapunov}
A \emph{mean-field Lyapunov function} of a PIN model  with limit drift $\ov D$ 
is a differentiable map $$V:\mc P(\mc A)\to\R_+\,,$$  such that
\be\label{eq:Lyapunov-mf22}\nabla V(\theta)\cdot  \ldrift(\theta)\leq 0\,,\qquad \forall \theta\in\mc P(\mc A)\,.\ee
\end{definition}

\begin{remark}
In the special case when the interaction pattern is a complete graph the considered PIN models are often referred to as a mean-field models. 
As mentioned in the Introduction, a well-known result in this context is Kurtz's theorem \cite{Kurtz:70} (see also \cite{DM:2010}). This result states that, if the limit drift $\ldrift(\theta)$ is Lipschitz-continuous on $\mc P(\mc A)$, then the type process $\Theta(nt)$ of a PIN model on a complete interaction pattern converges in probability as the order $n$ grows large to the solution of the mean-field ODE 
\be\label{ODE1}\dot\theta=\ldrift(\theta)\,,\ee
over bounded time horizons $t\in[0,T]$. 
Hence, by Definition \ref{def:meanfieldLyapunov}, a mean-field Lyapunov function $V(\theta)$ for a PIN model is a differentiable Lyapunov function for the mean-field ODE \eqref{ODE1}. 

Notice that, if $V(\theta)$ is a mean-field Lyapunov function, then the celebrated Barbashin-Krasovskii-LaSalle invariance principle for deterministic dynamical systems \cite[Theorem 4.4]{Khalill:02} implies that all solutions of the mean-field ODE \eqref{ODE1} converge, as $t$ grows large, to a subset of the set $\mc Z=\{\theta\in\mc P(\mc A):\,\nabla V(\theta)\cdot  \ldrift(\theta)= 0\}$ (more specifically, the invariance principle ensures convergence to the largest invariant subset of $\mc Z$). In the remainder of this section we shall present results with an analogous flavor for the type process $\Theta(t)$ of stationary stochastic PIN models. It is worth pointing out that our analysis will not rely on Kurtz's theorem and the Barbashin-Krasovskii-LaSalle invariance principle. In fact, our results will apply to PIN models with non-complete interaction patterns as well as discontinuous drift functions for which Kurtz's theorem does not apply. 
\end{remark}\medskip

\subsection{Asymptotically totally mixing networks}\label{sec:ATM-networks}
We now introduce a key mixing property for graphs.  
\begin{definition}\label{def:total-mixing}
The \emph{total mixing gap} of a nonempty graph $\mc G=(\mc V,\mc E)$ is the quantity
\be\label{W-graph}W_{\mc G}=\max\limits_{\begin{array}{l}\mc S, \mc U\subseteq \mc V\\ \mc S\cap\mc U =\emptyset\end{array}}
\left|\frac{|\mc E_{\mc S\mc U}|}{m}-\frac{|\mc S||\mc U|}{n(n-1)}\right|\,,\ee
where $$\mc E_{\mc S\mc U}=\{(s,u)\in\mc E\;|\; s\in \mc S,\, u\in\mc  U\}\,,$$
is the set of links pointing from nodes in $\mc S$ to nodes in $\mc U$, $n=|\mc V|$, and $m=|\mc E|$.\end{definition}\medskip

Clearly, for every nonempty graph $\mc G$, we have that $W_{\mc G}\ge0$. Moreover, since $|\mc E_{\mc S\mc U}|\le |\mc E|=m$ and $|\mc S||\mc U|\le n^2/4$ for every $\mc S\subseteq \mc V$ and $\mc U\subseteq\mc V$ such that $\mc S\cap\mc U=\emptyset$, we get that 
$$W_{\mc G}=\max\limits_{\mc S, \mc U}\left|\frac{|\mc E_{\mc S\mc U}|}{m}-\frac{|\mc S||\mc U|}{n(n-1)}\right|\le\max\left\{1,\frac{n}{4(n-1)}\right\}=1\,.$$
Hence, 
\be\label{eq:mixing-range-1}0\le W_{\mc G}\le 1\,,\ee
for every nonempty graph $\mc G$. 
On the other hand, if the graph $\mc G$ is undirected, then $|\mc E_{\mc S\mc U}|\le m/2$ for every $\mc S\subseteq \mc V$ and $\mc U\subseteq\mc V$ such that $\mc S\cap\mc U=\emptyset$, so that 
$$W_{\mc G}=\max\limits_{\mc S, \mc U}\left|\frac{|\mc E_{\mc S\mc U}|}{m}-\frac{|\mc S||\mc U|}{n(n-1)}\right|\le\max\left\{\frac12,\frac{n}{4(n-1)}\right\}=\frac12\,.$$ 
Hence, 
\be\label{eq:mixing-range-2}0\le W_{\mc G}\le 1/2\,,\ee
for every nonempty undirected graph $\mc G$. 
The following examples illustrate how the inequalities \eqref{eq:mixing-range-1} and \eqref{eq:mixing-range-2} are tight. 

\begin{example}\label{example:complete}
In the special case when $\mc G$ is the complete graph, we have $m=n(n-1)$ and $\mc E_{\mc S\mc U}=\mc S\times\mc U$ whenever $\mc S\cap\mc U=\emptyset$, so that $W_{\mc G}=0$. In fact, the converse is also true: if $W_{\mc G}=0$, then for every $s\ne u$ in $\mc V$ we have that  $$\left||\mc E_{\{s\}\{u\}}|-\frac{m}{n(n-1)}\right|\le m W_{\mc G}=0\,,$$
and since $|\mc E_{\{s\}\{u\}}|\in\{0,1\}$ and $m>0$ (because the graph is nonempty), it must be that $\mc E=\{(s,u):s\ne u \in\mc V\}$ and $m=n(n-1)$, i.e., $\mc G$ is the complete graph. \end{example}

\begin{example}\label{example:single-link}
For a graph $\mc G=(\mc V,\mc E)$ with a single directed link $\mc E=\{(s,u)\}$, we get that 
$$W_{\mc G}\ge\left||\mc E_{\{s\}\{u\}}|-\frac{m}{n(n-1)}\right|=1-\frac{1}{n(n-1)}\,,$$
that converges to $1$ as the order $n$ grows large.  \end{example}

\begin{example}\label{example:star}
{For the star graph with $n-1$ undirected links connecting a single node $s$ to all other nodes ${\cal U}:={\cal V}\setminus\{s\}$ we get 
\begin{align*}
W_{\mc G}&\ge\left|\frac{|\mc E_{\{s\}\,{\cal U}}|}{m}-\frac{|\{s\}|\ |{\cal U}|}{n(n-1)}\right|\\
&=\left|\frac{n-1}{2(n-1)}-\frac{n-1}{n(n-1)}\right|=\frac12-\frac1n\,,
\end{align*}
that converges to $1/2$ as the order $n$ grows large. 
}
 \end{example}\medskip

The following definition captures a fundamental asymptotic property of sequences of graphs. 

\begin{definition}\label{def:tfm}
A sequence of graphs $\mc G_{n}=(\mc V_{n}, \mc E_{n})$ is \emph{asymptotically totally mixing (ATM)} if 
$$\lim_{n\to+\infty}W_{\mc G_{n}}=0\,.$$
\end{definition}\medskip

Notice that a sequence $\mc K_n$ of complete graphs of increasing order is an ATM network since $W_{\mc K_n}=0$ for all $n\ge2$, as shown in Example \ref{example:complete}. In contrast, a sequence of star graphs $\mc S_n$ with increasing order $n$ is not an ATM network, since $W_{\mc S_n}\to1/2$ as $n$ grows large, as shown in Example \ref{example:star}.

In fact, an important example of ATM networks is provided by the Erd\"os-R\'enyi random graphs. These are defined as follows: for a positive integer $n$ and a parameter $p$ in $]0,1]$, let $\mc G(n,p)$ be the random undirected graph of order $n$ whereby pairs of distinct nodes are connected with probability $p$, independently from one another. 
Then, we have the following result.  

\begin{proposition}\label{prop:Erdos-Renyi} 
If $np\to+\infty$ as $n$ grows large, then the sequence of  Erd\"os-R\'enyi random graphs $\mc G(n,p)$ is an ATM network with probability $1$. 
\end{proposition}
\begin{IEEEproof}See Appendix  \ref{proof:prop:Erdos-Renyi}.\end{IEEEproof}\medskip


\subsection{Concentration of PIN models on ATM interaction networks}\label{sec:concentration-results}

We are now in a position to formally state the two main technical results of this paper concerning concentration properties of PIN models on ATM interaction networks.

%
%
%
%
%
%
%
%

\begin{theorem}\label{cor:ATM-general-2} 
Consider a PIN model with parameters $(\mc A, \pint, P, \phi)$ on an ATM interaction network $\mc G_{n}$. Let $\mu_{n}$ be a stationary distribution for all $n\ge2$. 
Denote by $\ldrift$ the limit drift and let $V:\mc P(\mc A)\to\R$ be a class-$\mc C^2$ mean-field Lyapunov function. 
Then
%
\be\label{mu-conc-mf22} \lim\limits_{n\to +\infty}\mu_{n} \{\xb\in\mc X:\,\nabla V(\theta(\xb))\cdot\ldrift(\theta(\xb))>-\delta\}=1\,,\ee
 for every $\delta>0$. 
\end{theorem}
\begin{IEEEproof}
See Section \ref{sec:proofs}.
\end{IEEEproof}
\medskip



As illustrated in the next section, there are cases when Theorem \ref{cor:ATM-general-2}  cannot be directly applied, e.g., since one cannot find a class-$\mc C^2$ mean-field Lyapunov function for the PIN model. In this case, we sharpen our technique by resorting to an approximate mean-field  Lyapunov function.
This result will be then applied to the SIRS model in Section \ref{sec:examples}.

{
\begin{definition}\label{def:approxmeanfieldLyapunov}
Consider a family of PIN models with parameters $(\mc A, \pint, P_\alpha, \phi_\alpha)_{\alpha\ge0}$, and let $\ldrift_\alpha(\theta)$, for $\alpha\ge0$, be their limit drift.  An \emph{approximate mean-field Lyapunov function} for such a family
is a family of class-$\mc C^2$ functions  
$$V_\alpha:\mc P(\mc A)\to\R\,,\qquad \alpha>0\,,$$
such that there exist two nonnegative-valued functions $h:\mc P(\mc A)\to\R_+$ and $\zeta:\R_+\to\R_+$ where $h(\theta)$ is continuous on $\mc P(\mc A)$ and
\be\label{limalpha}\lim_{\alpha\to 0}\zeta(\alpha)=0\,,\ee
such that
\be\label{eq:Valpha}
\nabla V_\alpha(\theta)\cdot  \ldrift_\alpha(\theta)\leq -h(\theta)+\zeta(\alpha)\,,\qquad \forall \theta\in\mc P(\mc A)\,.
\ee
\end{definition}
}


We can now state the following result, generalizing Theorem \ref{cor:ATM-general-2}. 

\begin{theorem}\label{coro-approx-ER} 
{
Consider a family of PIN models with parameters $(\mc A, \pint, P_\alpha, \phi_\alpha)_{\alpha\ge0}$ on an ATM interaction network $\mc G_{n}$ and an associated approximate mean-field Lyapunov function $V_\alpha$ as described in Definition \ref{def:approxmeanfieldLyapunov}.}
For every $\alpha> 0$ and $n\ge2$, let $\mu_n^{(\alpha)}$ be a stationary distribution. 
Then,
\be\label{mu-conc-mf2er} \lim_{\alpha\to 0}\liminf\limits_{n\to +\infty}\mu_n^{(\alpha)}\{\xb\in\mc X:\,h(\theta(\xb))<\delta\}=1\,,\ee
 for every $\delta>0$.
\end{theorem}
\begin{IEEEproof}See Section \ref{sec:proofs}.\end{IEEEproof}\medskip

\begin{remark}
Observe that, if a class-$\mc C^2$ function $V_0:\mc P(\mc A)\to\R_+$ existed such that both $V_\alpha(\theta)\stackrel{\alpha\to0}{\longrightarrow}V_0(\theta)$ and $\nabla V_\alpha(\theta)\stackrel{\alpha\to0}{\longrightarrow}\nabla V_0(\theta)$ uniformly on $\mc P(\mc A)$, then taking the limit of both sides of \eqref{eq:Valpha} would  yield $$\nabla V_0(\theta)\cdot  \ldrift_0(\theta)\leq -h(\theta)\le0\,,\qquad \forall \theta\in\mc P(\mc A)\,.$$
{In this case, $V_0(\theta)$ would be a mean-field Lyapunov function for the PIN model with parameters $(\mc A, \pint, P_0, \phi_0)$ and Theorem \ref{cor:ATM-general-2} could be applied directly. However, there are families of PIN models such as the SIRS model described in Example \ref{ex:SIRS} of Section \ref{sec:nonbinary}, which admit approximate mean-field Lyapunov function $V_{\alpha}$ 
that does not even admits a limit
as $\alpha$ vanishes.
Theorem \ref{coro-approx-ER} proves useful in addressing exactly these cases.}
\end{remark}

\section{Examples of applications}\label{sec:examples}
In this section, we introduce some classes of PIN models and apply the concentration results of the previous section.

\subsection{Forgetful PIN models}\label{sec:forgetful}
We first consider the special case of PIN models where the entries of the interaction transition tensor $\phi_{ij}(\ell)$ do not depend on the current state $i$. Precisely, we assume that there exists a row-stochastic matrix $R$ such that
$$\phi_{ij}(\ell)=R_{\ell j}\,,\qquad\forall i,j,\ell\in\mc A\,.$$
Notice that the limit drift for these models has the form
\be\label{drift-forgetful}\ldrift(\theta)=(S^\top-I)\theta\,,\qquad \forall\theta\in\mc P(\mc A)\,,\ee
where
\be\label{S-def}S=(\pmut) P+\pint R\,,\ee
is row-stochastic, i.e., the matrix associated with the linear map $\ldrift(\theta)$ is the transpose of a Laplacian matrix.
{ These models are referred to as forgetful PIN models and arise in evolutionary game dynamics \cite{Lieberman.ea:05,Nowak:06}. }

Assume that the row-stochastic matrix $S=(\pmut) P+\pint R$ is irreducible. It then follows from the Perron-Frobenius theorem that $S$ admits a unique invariant probability vector $\pi=S^\top\pi$ in $\mc P(\mc A)$. 
For such models over ATM networks, Theorem \ref{cor:ATM-general-2} yields a concentration of the type process on the vector  $\pi$. Precisely, we have the following result.

\begin{proposition} \label{forgetful-asymp} 
Consider a forgetful PIN model over an ATM network  $\mc G_{n}$. 
Assume that the stochastic matrix $S$ in \eqref{S-def} is irreducible and let $\pi=S^\top\pi$ in $\mc P(\mc A)$ be its unique invariant probability vector. 
Let  $\mu_{n}$ be a stationary distribution for all $n\ge2$. Then, 
%
%
%
$$\lim\limits_{n\to +\infty}\mu_{n}\{\xb\in\mc X\;|\;\|\theta(\xb)-\pi\|_2<\delta\}=1\,.$$
for every $\delta>0$. 

%
\end{proposition}

\begin{IEEEproof} 
We first construct a quadratic function $V(\theta)$ that is positive definite with respect to $\pi$ and such that \eqref{eq:Lyapunov-mf22} holds true. 
To this aim, we consider the Laplacian matrix $L:=I-S^\top$ and we recall some basic well-known properties:
\begin{itemize}\item all eigenvalues $\lambda$ of $L$ are such that $\Re(\lambda)\geq 0$;
\item the only eigenvalue on the imaginary axis is $0$ and this is simple with left and right eigenvectors being, respectively, $\1$ and $\pi$.
\end{itemize}
As a consequence, the spectrurm of the matrix $$A=L+\pi\1^\top$$ lies in the open half-plane $\Re(\lambda)> 0$ and, consequently, the Lyapunov equation
\be\label{quadratic-Lyapunov}\Pi A+A^\top \Pi=I\,,\ee
admits a solution $\Pi$ in $\R^{\mc A\times\mc A}$ that is symmetric and positive definite. Consider now the function $V:\mc P(\mc A)\to\R$, defined as
\be\label{quadratic-V}V(\theta)=(\theta-\pi)^\top \Pi(\theta-\pi)\,,\qquad\forall\theta\in\mc P(\mc A)\,.\ee 
Clearly, $V(\theta)$ is of class $\mc C^2$. Moreover, it follows from \eqref{drift-forgetful} that
{$$
\ba{rcl}\nabla V(\theta)\cdot \ldrift(\theta)
&=&-2(\theta-\pi)^\top \Pi L\theta\\[5pt]
&=&-2(\theta-\pi)^\top \Pi(L+\pi\1^\top)(\theta-\pi)\\[5pt]
&=&-(\theta-\pi)^\top\left(\Pi A+A^\top \Pi\right)(\theta-\pi)\\[5pt] 
&=&-||\theta-\pi||_2^2\le 0\,.
\ea$$
}
This shows that \eqref{eq:Lyapunov-mf22} is satisfied, so that $V(\theta)$ is a mean-field Lyapunov function for the considered PIN model. 
The result now follows from Theorem \ref{cor:ATM-general-2}.
\end{IEEEproof}\medskip
%
%
%
%
%
%
%
%
%

{ 
{Explicit examples of forgetful PIN models come from evolutionary game theory, as for instance the following best response learning model. }
\begin{example}
Consider a $2$-player symmetric game with set of actions $\mc A$ and payoff matrix $U$ in $\R^{\mc A\times\mc A}$ (i.e., $U_{ij}$ is the payoff obtained by the first player when she plays $i$ and her opponent plays $j$). We assume the agents to interact in a pairwise fashion and to choose best response actions. 
Precisely, 
the jump interaction probabilities are given by
$$R_{\ell j}=\left\{\begin{array}{ll} |\mc B(\ell)|^{-1}\quad &\hbox{if}\; j\in \mc B(\ell)\\ 0 \quad &\hbox{otherwise}\end{array}\right.$$
where $\mc B(\ell)=\argmax\limits_{k\in\mc A}U_{k\ell}$
is the best response set. 
\end{example}
}

\subsection{Binary PIN models}\label{sec:binary}

We now analyze in detail general PIN models with binary state space $\mc A=\{0,1\}$. Notice that in this case seven parameters are sufficient to specify the model: the interaction probability $\rho$, two mutation probabilities $P_{01}$ and $P_{10}$, and the four elements of the interaction transition tensor $ \phi_{01}(0)$, $ \phi_{01}(1)$, $\phi_{10}(0)$, and $\phi_{10}(1)$.
In particular, the limit drift takes the following form
$$\ba{rcl}
\ldrift_1(\theta)\!\!\!&\!\!\!\!=\!\!\!\!&\!\!\!(\pmut)\left(P_{01}\theta_0-P_{10}\theta_1\right)\\[5pt]
&&\!\!\!+\pint
\left(\phi_{01}(0)\theta_0^2+(\phi_{01}(1)\!-\!\phi_{10}(0))\theta_0\theta_1-\phi_{10}(1)\theta_1^2\right).
\ea$$

By substituting $\theta_0=1-\theta_1$, we get that $${\ldrift}_1(\theta)=p(\theta_1)$$ where $p$ is the quadratic polynomial 
\be\label{poly}p(z)=a_0+a_1z+a_2z^2\ee
with coefficients 
\be\label{coeff}\begin{array}{rcl}a_0\!\!&\!\!\!=\!\!\!& \!\!(\pmut) P_{01}+\pint\phi_{01}(0)\\[5pt]
a_1\!\!&\!\!\!=\!\!\!&\!\!-(\pmut)\left(P_{01}+P_{10}]+\pint[-2\phi_{01}(0)+\phi_{01}(1)-\phi_{10}(0)\right)\\[5pt]
a_2\!\!&\!\!\!=\!\!\!&\!\! \pint\left(\phi_{01}(0)-\phi_{01}(1)+\phi_{10}(0)-\phi_{10}(1)\right)\,.
\end{array}
\ee

Notice that 
$$\begin{array}{l}p(0)=(\pmut) P_{01}+\pint\phi_{01}(0)\geq 0\,,\\[5pt] p(1)=-(\pmut) P_{10}-\pint\phi_{10}(1)\leq 0\,,\end{array}$$
so that there always exists at least one zero of $p(z)$ in the interval $[0,1]$. When $p(0)>0$ and $p(1)<0$ (this happens for instance when the two mutation terms $P_{01}$ and $P_{10}$ are both nonzero) such zero is unique and is denoted by $z^*$. 
{The next result shows that, under suitable assumptions, $z^*$ is exactly where concentration of the component $1$ of the type process takes place.}

\begin{proposition} \label{binary-asymp} 
Consider a binary PIN model over an ATM interaction network $\mc G_n$. 
Let $\mu_{n}$ be a stationary distribution of the PIN model for all $n\ge2$.
If the polynomial $p(z)$ defined in \eqref{poly} admits a unique zero $z^*$ in $[0,1]$, 
then $$\lim\limits_{n\to +\infty}\mu_{n}\{\xb\in\mc X\;|\; |\theta_1(\xb)-z^*|<\delta\}=1\,,$$
for every $\delta>0$.
\end{proposition}
\begin{IEEEproof}
%
%
%
Consider $V:\mc P(\mc A)\to\R$ defined by
$$V(\theta)=-\left(a_0\theta_1+\frac{1}{2}a_1\theta_1^2+\frac{1}{3} a_2\theta_1^3\right)\,,\qquad \forall\theta\in\mc P(\mc A)\,.$$
Notice that 
$$\nabla V(\theta)\cdot\ldrift(\theta)=-p(\theta_1)^2\,.$$
Therefore,
$$\{\theta\in\mc P(\mc A)\;|\; \nabla V(\theta)\cdot\ldrift(\theta)=0\}=\{(1-z^*, z^*)\}$$
and the result now follows from Theorem \ref{cor:ATM-general-2}.
\end{IEEEproof}\medskip

%


\begin{example}[SIS epidemic models]\label{ex:SIS}
In the SIS epidemic model,  the two states $0$ and $1$ represent, respectively, that an individual is either susceptible or infected. In this model, $\phi_{01}(1)$ is the only nonzero term of the interaction transition tensor.
We interpret $b:=\pint\phi_{01}(1)$ as the contagion probability, $c:=(\pmut) P_{10}$ as the recovery probability, and $\alpha:=(\pmut) P_{01}$ as the spontaneous infection probability. 
%
%
%
%
%
%
The polynomial in \eqref{poly} becomes
$$p(z)=\alpha+(b-c-\alpha)z-bz^2\,.$$
 {
 In the standard case in which $\alpha=0$ the PIN model is not ergodic and the only stationary distribution consists in all the agents being susceptible that is the absorbing state of the Markov chain. From now on,  we assume that $\alpha>0$ together with} 
$b>0$ and $c>0$. We indicate with $z^*(\alpha)$ the unique zero of $p(z)$ in $[0,1]$ as a function of $\alpha$ and we notice that 
%
%
%
\be\label{SIS-equilibrium}\lim\limits_{\alpha \to 0}z^*(\alpha)=z^{*}=\max\{1-c/b,0\}\,.\ee
{
Proposition \ref{binary-asymp} implies that the stationary distribution of the associated PIN model
over an ATM interaction network $\mc G_n$ concentrates around the point $z^*(\alpha)$. 
Considering the limit also in $\alpha$ we further get
$$\lim\limits_{\alpha\to 0^+}\lim\limits_{n\to +\infty}\mu_n^{(\alpha)}\{\xb\;|\;|\theta_1(\xb)-z^*|<\delta\}=1\,,
$$
for every $\delta >0$.
}
We notice that $z^*$ coincides with the asymptotically stable equilibrium of the corresponding mean field ODE \eqref{ODE1}. {In particular, this shows that the well known bifurcation phenomenon according which for $c>b$ the epidemics dies out while for $c<b$ it becomes endemic at level $1-c/b$, holds true not only for complete interaction patterns, but also for general ATM interaction networks, as for instance Erd\"os-R\'enyi random graphs.}
\end{example}\medskip

Other popular examples of PIN models falling into this family are reported below.

\begin{example}[Voter/Anti-voter models]\label{ex:voter} In binary opinion dynamics models, the two states $0$ and $1$ represent two different opinions. In the \emph{noisy voter model} \cite{Granovsky.Madras:95,Carro.ea:16}, agents modify their opinion in pairwise interactions by copying the opinion of the agent they meet or because of a mutation (equal for the two states). This is represented by the following choices of the parameters:
$$\phi_{01}(1)=\phi_{10}(0)=1\,,\qquad\phi_{01}(0)=\phi_{10}(1)=0\,,$$ 
and $P_{01}=P_{10}=1$.  
In this case, we have 
\begin{align*}
 \ldrift_1(\theta)&=(\pmut)(\theta_0-\theta_1).
\end{align*}

In the \emph{anti-voter} model \cite{Donnelly.Welsh:84,Liggettbook}  no mutation is present, namely $\pint=1$, and agents take the opposite opinion of the agent they meet. Precisely,
$$\phi_{01}(1)=\phi_{10}(0)=0\,,\qquad\phi_{01}(0)=\phi_{10}(1)=1\,.$$ 
In this case, we have 
\begin{align*}
 \ldrift_1(\theta)&=\theta_0^2-\theta_1^2.
\end{align*}
{Notice that in both examples we have $z^*=1/2$.}
\end{example}\medskip

\subsection{Non-binary models}\label{sec:nonbinary}
Finally, we present an example that does not fit in any of the two categories  above.

\begin{example}[SIRS epidemic model]\label{ex:SIRS}
In the SIRS epidemic model we have $\mc A=\{0,1,2\}$ where $\xb_v=0$ means the individual $v$ is susceptible, $\xb_v=1$ means that the individual $v$ is infected and $\xb_v=2$ means that the individual $v$ is recovered. The non-zero off-diagonal terms in $P$ are $P_{01}$, $P_{12}$, and $P_{20}$, the conditional probability of, respectively, spontaneous infection, spontaneous recovery, and spontaneous return to susceptibility. As in the SIS model, the only off diagonal term of $\phi(\ell)$ is $\phi_{01}(1)$ the conditional probability of contagion transmission.
It is convenient to set the following variables:
$$b:=\pint\phi_{01}(1)\,,\quad c:=(\pmut) P_{12}\,,\quad  d:=(\pmut) P_{20}\,,$$ 
$$\alpha:=(\pmut) P_{01}\,.$$ 
We assume that $\alpha >0$ and moreover that $b>c>0$ and $d>0$.
We denote by ${\ldrift}_{\alpha}(\theta)$ its limit drift. It can be seen that
\begin{align*}
{\ldrift}_{1,\alpha}(\theta)&=b(1-\theta_1-\theta_2)\theta_1-c\theta_1+\alpha(1-\theta_1-\theta_2)\\[5pt]
&=(\theta_1+\alpha/b)[b(1-\theta_1-\theta_2)-c]+\alpha c/b\\[5pt]
&= -b(\theta_1+\alpha/b)[(\theta_1-z_1^*)+(\theta_2-z_2^*)]+\alpha c/b\,,\\[5pt]
{\ldrift}_{2,\alpha}(\theta)&=c\theta_1-d\theta_2=c(\theta_1-z_1^*)-d(\theta_2-z_2^*)\,,
\end{align*}
where
\be\label{z-SIRS}z_1^*=\frac{d(b-c)}{b(d+c)},\quad z_2^*=\frac{c(b-c)}{b(d+c)}\,.\ee

Now, consider the family of functions 
$$V_\alpha(\theta_1,\theta_2):=\theta_1-(z^*_1+\alpha/b)\ln(\theta_1+\alpha/b)+\frac{b}{2c}(\theta_2-z^*_2)^2\ .$$
Notice that these function are of class $\mc C^2$ on $\mc P(\mc A)$ for $\alpha>0$ and 
 have partial derivatives
\be\label{partial-SIRS}\frac{\partial V_\alpha}{\partial\theta_1}(\theta)=\frac{\theta_1-z_1^*}{\theta_1+\alpha/b},\qquad \frac{\partial V_\alpha}{\partial\theta_2}(\theta)=\frac{b}{c}(\theta_2-z_2^*)\,.\ee
Straightforward computations now yield
$$\ba{rcl}
&&\ds\nabla V_\alpha(\theta)\cdot {\ldrift}_{\alpha}(\theta)\\
&&\qquad =\ds-b\left(\theta_1-z_1^*\right)^2-\frac{bd}{c}\left(\theta_2-z_2^*\right)^2+\frac{\alpha c}{b}\frac{\theta_1-z_1^*}{\theta_1+\frac{\alpha}{b}}\\[7pt]
&&\qquad \le\ds-b\left(\theta_1-z_1^*\right)^2-\frac{bd}{c}\left(\theta_2-z_2^*\right)^2+\ds \frac{\alpha c}{b}
\,,
\ea$$
where the last inequality follows form the fact that the function $$g(z)=\frac{z-z_1^*}{z+\alpha/b}$$ is increasing in $z\in [0,1]$ and hence $g(z)\le 1$. 

We are then in a position to apply Theorem \ref{coro-approx-ER} by taking 
{$$h(\theta)=\ds-b\left(\theta_1-z_1^*\right)^2-\frac{bd}{c}\left(\theta_2-z_2^*\right)^2,\quad \zeta(\alpha)=\alpha c/b\,,$$} 
and to conclude that \eqref{mu-conc-mf2er} holds true. 
This essentially says that the invariant measure $\mu_n^{(\alpha)}$ converges, as $n$ grows large and as $\alpha$ tends to zero, to a probability measure whose projection by $\theta(\cdot)$ coincides with the Dirac delta distribution centered in $\theta^*=(1-z_1^*-z_2^*, z_1^*, z_2^*)$,
where $z_1^*$ and $z_2^*$ are defined in  \eqref{z-SIRS}.

\end{example}
\section{Proofs of the main concentration results}\label{sec:main-concentration}

In this section, we present all technical results and, in particular, we prove Theorems \ref{cor:ATM-general-2} and \ref{coro-approx-ER}. 
We first derive general approximation results 
that are valid for any PIN model on arbitrary interaction patterns. We then restrict to ATM interaction networks and we establish the two large scale concentration results.

\subsection{Mean drift}\label{sec:mean-drift}

Consider a PIN model with interaction pattern $\mc G=(\mc V,\mc E)$ and parameters $(\mc A,\pint, P,\phi)$, and let $\mc X$ be its configuration space. 
A key quantity in the analysis of the PIN model is its \emph{mean drift}, defined as the function $D:{\mc X}\to \R^{\mc A}$ mapping a configuration $\xb$ in $\mc X$ into the vector
\begin{equation}\label{eq:drift}
D(\xb):=n\E\l[\theta(X(t+1))-\theta(X(t))|X(t)=\xb\r]\,.
\end{equation}
Entries $D_i(\mb x)$ represent the conditional expected variation of the number of agents in a given state $i$, when the current configuration of the PIN model is $X(t)=\xb$. 
Notice that, since $\sum_{i\in\mc A}\theta_i(\xb)=1$ for every configuration $\xb$ in $\mc X$, we have that 
$$\ba{rcl}\ds0&=&\ds n\E\l[\sum_{i\in\mc A}\left(\theta_i(X(t+1))-\theta_i(X(t))\right)|X(t)=\xb\r]\\&=&\ds\sum_{i\in\mc A}D_i(\xb)\,,\ea$$
i.e., $D(\xb)$ is always a zero-sum vector. 

It also proves useful to introduce the notion of \emph{boundary} of a configuration $\xb$ in $\mc X$, defined as the vector $\xi(\xb)$ of the empirical frequencies of the pairs of states in $\xb$ that are connected by links of $\mc G$. Formally, let $\xi:\mc X\to\mc P(\mc A\times\mc A)$ be the function mapping  a configuration $\xb$ in $\mc X$ to $\xi(\xb)$ in $\mc P(\mc A\times\mc A)$ with entries 
\be\label{eq:boundary}\xi_{ij}(\xb):=\frac1{m}{|\{(u,v)\in{\mc E}\,|\,\xb_u=i,\  \xb_v=j\}|}\,,\qquad i,j\in\mc A\,,\ee
where we recall that $m=|\mc E|$ is the number of edges of the interaction pattern $\mc G$. 
Clearly, if $\mc G$ is undirected, then $\xi_{ij}(\xb)=\xi_{ji}(\xb)$ for every two states $i$ and $j$ in $\mc A$ and every configuration $\xb$ in $\mc X$. 

Now, we introduce the linear operator $\mc Q:\mc P(\mc A\times\mc A)\to\R^{\mc A}$ mapping a boundary $\xi$ in $\mc P(\mc A\times\mc A)$ into a zero-sum vector $\mc Q(\xi)$ in 
$\R^{\mc A}$ with entries 
\be\label{operQ}\left(\mc Q(\xi)\right)_i:=\sum\limits_{\ell\in\mc A}\sum\limits_{j\in\mc A}\xi_{j\ell}\phi_{ji}(\ell)-\sum\limits_{\ell\in\mc A}\xi_{i\ell}   \,,\qquad i\in\mc A\,.\ee
Then, we have the following result. 

\begin{lemma}\label{lemma:driftrep} The mean drift of a PIN model with interaction network $\mc G=(\mc V,\mc E)$ and parameters $(\mc A,\pint, P,\phi)$ satisfies
\be\label{eq-Dgeneral} D(\xb)=(\pmut)(P^\top-I)\theta(\xb)+\pint\mc Q(\xi(\xb))\,\ee 
for every configuration $\bf x$ in $\mc X$. 
\end{lemma}
\begin{IEEEproof}
Since the configuration $X(t)$ of a PIN model can change at most one of its entries at a time, we have that either 
$$\theta(X(t+1))-\theta(X(t))=0\,,$$ 
(when no change occurs) or 
$$\theta(X(t+1))-\theta(X(t))=n^{-1}(\delta^j-\delta^i)\,,$$
for some $i\neq j$ in $\mc A$ (when an agent modifies its state from $i$ to $j$). Call $E_{ij}$ the latter event and notice that, because of the update mechanism, we have that
$$\P(E_{ij}\,|\, X(t)=\xb)=(\pmut)\theta_i(\xb)P_{ij}+\pint\sum_{\ell\in{\mc A}}\xi_{i\ell}(\xb)\phi_{ij}(\ell)\,.$$
Thus,
$$\begin{array}{rcl}D_i(\bf x)&=&\ds\sum\limits_{j\neq i}\P(E_{ji}\,|\, X(t)=\xb)-\sum\limits_{j\neq i}\P(E_{ij}\,|\, X(t)=\xb)\\[10pt]
&=&\ds(\pmut)\sum\limits_{j\neq i}\theta_j(\xb)P_{ji}+\pint{ \sum\limits_{\ell\in\mc A}\sum\limits_{j\neq i}\xi_{j\ell}(\xb)\phi_{ji}(\ell)}\\[5pt]
&&\ds- (\pmut)\sum\limits_{j\neq i}\theta_i(\xb)P_{ij}-\pint{ \sum\limits_{\ell\in\mc A}\sum\limits_{j\neq i}\xi_{i\ell}(\xb)\phi_{ij}(\ell)}\\[10pt]
&=&(\pmut)\left(P^\top\theta(\xb)-\theta(\xb)\right)_i+\pint\left(\mc Q(\xi(\xb))\right)_i\,,
\end{array}$$
which yields the result.
\end{IEEEproof}\medskip

Lemma \ref{lemma:driftrep} states that the mean drift $D(\xb)$ of a PIN model depends on the configuration $\xb$ through both its type $\theta(\xb)$ and its boundary $\xi(\xb)$. This is a consequence of the particular structure of PIN models that allow for both individual mutations and pairwise interactions on the graph $\mc G$, but not, e.g., on higher order interactions. 

Below, we discuss in general how the mean drift and the limit drift are related.
We start with the following result. 

\begin{lemma}\label{lemma:estimate}
Consider a PIN model with interaction pattern $\mc G=(\mc V,\mc E)$ and parameters $(\mc A,\pint, P,\phi)$.
Let $D(\xb)$ be its mean drift and  $\ldrift(\theta)$ be its limit drift. 
Then, 
$$||D(\xb)- \ldrift(\theta(\xb))||_1\le2\pint||\xi(\xb)-\theta(\xb)\theta(\xb)^\top||_1\,,$$
for every configuration $\xb$ in $\mc X$. 
\end{lemma}
\begin{IEEEproof} 
For a given configuration $\xb$ in $\mc X$, write $\theta$ for $\theta(\xb)$ and $\xi$ for $\xi(\xb)$. Then, 
$$\ba{rcl}
||D(\xb)- \ldrift(\theta)||_1
&=&\pint ||{\mc Q}(\xi)-{\mc Q}(\theta\theta^\top)||_1\\[10pt]
&=& \ds\pint\sum_{i\in\mc A}|(\mc Q(\xi-\theta\theta^\top)_i|\\[10pt]
&\le& \ds\pint\sum_{i\in\mc A}\sum_{j\in\mc A}\sum_{\ell\in\mc A}|\xi_{j\ell}-\theta_j\theta_{\ell}|\phi_{ji}(\ell)\\[10pt]
&&\ds+\pint\sum_{i\in\mc A}\sum_{\ell\in\mc A}|\xi_{i\ell}-\theta_i\theta_{\ell}|\\[10pt]
&=& 2\pint||\xi-\theta\theta^\top||_1\,, 
\ea$$
thus proving the claim. 
\end{IEEEproof}\medskip

Lemma \ref{lemma:estimate} provides an upper bound on the $1$-norm distance between the mean drift $D(\xb)$ in a configuration $\xb$ and the corresponding limit drift $\ldrift(\theta(\xb))$ of a PIN model. This is no more than twice the $l_1$-distance between the boundary $\xi(\xb)$ of $\xb$ and the product distribution $\theta(\xb)\theta(\xb)^\top$. It is then clear that the limit drift  $\ldrift(\theta(\xb))$ provides a good approximation of the mean drift $D(\xb)$ when the boundary $\xi(\xb)$  is close to $\theta(\xb)\theta(\xb)^\top$ for every configuration $\xb$ in $\mc X$. 

A special case when this occurs is when the interaction pattern $\mc G$ is the complete graph. First notice that, in this case, the boundary $\xi(\xb)$ of a configuration $\xb$ is completely determined by its type $\theta(\xb)$. Specifically, for a complete graph with $n$ nodes, we have that $\xi(\xb)=\xi^{(n)}(\theta(\xb))$, where
\be\label{boundary-mf}\xi_{ij}^{(n)}(\theta):= \frac{n\theta_i\theta_j}{n-1}-\frac{\theta_i\delta^i_j}{n-1}=\theta_i\theta_j+\frac{\theta_i(\theta_j-\delta^i_j)}{n-1}\,.
\ee
As a consequence, also  the mean drift  $D(\xb)$ depends on the type $\theta(\xb)$  only: this is consistent with the observation made in Remark \ref{remark:complete} that, for complete interaction networks, the type $\theta(X(t))$ is itself a Markov chain.  Lemma \ref{lemma:estimate} and expression \eqref{boundary-mf} imply that, for complete interaction networks 
\be\label{eq:TV-1}||D(\xb)-\ldrift(\theta(\xb))||_{1}\le\frac{4\pint}{n-1}\,,\qquad\forall\xb\in\mc X\,,\ee
so that the mean drift and the limit drift coincide in the large-scale limit.

The following result can then be interpreted as a generalization of \eqref{eq:TV-1}. 
\begin{lemma}\label{lemma:1-norm}
For every configuration $\xb$ in $\mc X$
\be\label{eq:TV-2}||D(\xb)- \ldrift(\theta(\xb))||_{1}\le 2\pint|\mc A|^2W_{\mc G}+\frac{4\pint}{n-1}\,.\ee
\end{lemma}
\begin{IEEEproof}
{First, it follows from definitions \eqref{W-graph}, \eqref{eq:boundary}, and \eqref{boundary-mf}
that}
\be\label{Wdef}W_{\mc G}=\max_{\xb\in\mc X}\left|\left|\xi(\xb)-\xi^{(n)}(\theta(\xb))\right|\right|_{\infty}\,.\ee
Then, for every configuration $\xb$ in $\mc X$, we have that 
\be\label{eqlemma3}\ba{rcl}
\ds||D(\xb)- \ldrift(\theta(\xb))||_{1}
&\le&\ds2\pint||\xi(\xb)-\theta(\xb)\theta(\xb)^\top||_1\\[7pt]
&\le&\ds2\pint||\xi(\xb)-\xi^{(n)}(\theta(\xb))||_{1}\\[5pt]
&&+\ds2\pint||\xi^{(n)}(\theta(\xb))-\theta(\xb)\theta(\xb)^\top||_1\\[7pt]
&\le&\ds 2\pint|\mc A|^2W_{\mc G}+\frac {4\pint}{n-1}\,,
\ea\ee
where the first inequality follows from Lemma \ref{lemma:estimate},   the second  inequality is a consequence of the triangle inequality,  and the last inequality is implied by \eqref{Wdef} and \eqref{boundary-mf}. 
\end{IEEEproof}\medskip

Lemma \ref{lemma:1-norm} states that the mean drift $D(\xb)$ is close to the limit drift $\ov D(\theta(\xb))$ when the order $n$ is large and the total mixing gap of the interaction pattern $W_{\mc G}$ is small.

\subsection{Concentration results for finite networks}\label{sec:concentration-finite}
Below, we shall use the notation $\E_{\mu}$ to indicate the expectation with respect to a stationary distribution $\mu$.
The following technical result, that is a simple consequence of stationarity, will prove to be very useful in our subsequent derivations. 

\begin{lemma}\label{lemma1} Consider a PIN model and let $\mu$ in $\mc P(\mc X)$ be one of its stationary distributions. Moreover, let $D(\xb)$ be its mean drift. 
Then, for every class-$\mc C^2$ function $V:\mc P(\mc A)\to\R$, we have that
$$\left|\E_\mu\l[\nabla V(\theta(X(t)))\cdot D(X(t))\r]\right|\le\frac{K}{n}\,,$$
where $K$ is a non-negative constant only depending on $V$.
\end{lemma}

\begin{IEEEproof}See Appendix \ref{sec:proof:lemma1}. 
\end{IEEEproof}\medskip

Lemma \ref{lemma1} allows us to prove the following result for PIN models in stationarity.

\begin{proposition}\label{prop-approximate-Lyapunov1}  
Consider a PIN model. Let $\mu$ in $\mc P(\mc X)$ be one of its stationary distributions and let $D(\xb)$ be its mean drift. 
Assume there exists a class-$\mc C^2$ function $V:\mc P(\mc A)\to\R$ such that
\be\label{eq:Lyapunov0}\nabla V(\theta(\xb))\cdot  D(\xb)\leq -F(\xb)+\epsilon\,,\qquad \forall \xb\in\mc X\,,\ee
for some $F:\mc X \to\R_{+}$ and $\epsilon \ge 0$.
Then, there exists a constant $K>0$ only depending on $V$ such that
\be\label{mu-conc} \mu \l(\l\{\xb\in\mc X\,|\,F(\xb)<\delta \r\}\r)\ge1-\frac{K}{n\delta}-\frac{\epsilon}{\delta}\,,\ee
for every $\delta >0$.
\end{proposition}

\begin{IEEEproof} 
It follows from 
Lemma \ref{lemma1} and inequality \eqref{eq:Lyapunov0} that
$$\ba{rcl}
\!\!\!\ds\E_\mu\l[F(X(t))\r]\!\!
&\!\!\!\!\le\!\!\!\!&\!\!\ds|\E_\mu\l[\nabla V(\theta(X(t)))\cdot D(X(t))\r]|\\[7pt]
&&\!\!\ds+\E_\mu\l[\nabla V(\theta(X(t)))\cdot D(X(t))+F(X(t))\r]\\[7pt]
&\!\!\!\!\le\!\!\!\!&\!\!
\ds\frac{K}{n}+\epsilon\,.
\ea$$
Now, by applying the Markov inequality to the nonnegative random variable $F(X(t))$, we get
$$\mu \l(\l\{X(t)\in\mc X\,|\,F(X(t))\ge \delta\r\}\r)\leq 
\frac{\E_\mu [F(X(t))]}{\delta}\leq\frac{K}{n\delta}+\frac{\epsilon}{\delta}\,,$$
thus proving the claim.
\end{IEEEproof}\medskip

{
\begin{remark}
Proposition \ref{prop-approximate-Lyapunov1} implies that {when inequality \eqref{eq:Lyapunov0} holds with $\epsilon =0$} and $n$ grows large the stationary distributions tends to concentrate where $F(\xb)$ is close to zero.
In particular when 
\be\label{V-D}\nabla V(\theta(\xb))\cdot  D(\xb)\le 0\,,\qquad \forall \xb\in\mc X\,,\ee
we have that the stationary distributions tend to concentrate where $\nabla V(\theta(\xb))\cdot  D(\xb)$ is close to zero. 
Inequality \eqref{V-D} states that $V(\theta(X(t)))$ is non-increasing in expectation for the process $X(t)$ in stationarity. Therefore, in this case, we can interpret $V$ as a Lyapunov function for  $X(t)$.  Notice that the mean drift $D(\xb)$ depends not only on the parameters $(\mc A,\pint, P,\phi)$ of the PIN model but also on the interaction pattern $\mc G$. Hence, in general, the search for a Lyapunov function $V$ satisfying \eqref{V-D} is hard since it has to consider also the interaction pattern $\mc G$. This essentially limits the applicability of Proposition \ref{prop-approximate-Lyapunov1}  for $\eps=0$ to the case of a complete interaction pattern. 

\end{remark}
\medskip
}

In the following, we shall  focus on the application of Proposition \ref{prop-approximate-Lyapunov1} with $\eps>0$, using class-$\mc C^2$ functions $V:\mc P(\mc A)\to\R$ that in general do not satisfy \eqref{V-D}, but instead are such that
\be\label{eq:Lyapunov-mf*}\nabla V(\theta)\cdot  \ldrift(\theta)\leq \zeta\,,\qquad \forall \theta\in\mc P(\mc A)\,,\ee
for some nonnegative constant $\zeta\ge0$, where we recall the $\ldrift(\theta)$ is the limit mean drift as defined in \eqref{drift-mf}.   
Notice that the limit mean drift $\ldrift(\theta)$ does not depend on the interaction pattern $\mc G$ of the considered PIN model, but only on its parameters. 
We are now ready to formulate our main concentration result. 
Given a class-$\mc C^2$ function $V:\mc P(\mc A)\to\R$ and constant $\delta>0$,
define
\be\label{Xdelta}\mc X_{\delta}=\{\xb\in\mc X:\,\nabla V(\theta(\xb))\cdot\ldrift(\theta(\xb))>-\delta\}\,.\ee

\begin{theorem}\label{theo:general} 
Consider a PIN model with parameters $(\mc A,\pint, P,\phi)$ and interaction pattern $\mc G$.  Let $\mu$ in $\mc P(\mc X)$ be one of its stationary distributions and let $\ldrift(\theta)$ be its limit mean drift. 
Assume that there exists a class-$\mc C^2$ function $V:\mc P(\mc A)\to\R_+$ such that
\eqref{eq:Lyapunov-mf*}
holds for some $\zeta\ge0$. 
Then, there exists a constant $C>0$ only depending on $V$ such that, for every $\delta >\zeta$ it holds
{
\be\label{mu-conc-mf} \mu \l(\mc X_{\delta-\zeta}\r)\ge1-\frac{C}{n\delta}-2 |\mc A|^2||\nabla V||_{\infty}\frac{W_{\mc G}}{\delta}-\frac{\zeta}{\delta}\,,\ee
}
where $||\nabla V||_{\infty}$ is the infinity norm of the function $\nabla V(\theta)$.
\end{theorem}

\begin{IEEEproof}
We can write
\be\label{estim-Lyap}\nabla V(\theta(\xb))\cdot  D(\xb)\leq \nabla V(\theta(\xb))\cdot  \ldrift(\theta(\xb))+\ov\epsilon\ee
where
\be\label{epsn}\ba{rcl}\ov\epsilon
&:=&\ds\max_{\xb\in\mc X}|\nabla V(\theta(\xb))\cdot  ( D(\xb)- \ldrift(\theta(\xb)))|\\[10pt]
&\le&\ds||\nabla V||_{\infty}\max_{\xb\in\mc X}||D(\xb)- \ldrift(\theta(\xb))||_{1}\\[10pt]
&\le&\ds||\nabla V||_{\infty}\left(2  |\mc A|^2W_{\mc G}+\frac{4}{n-1}\right)\,,
\ea\ee
and where the second inequality follows from Lemma \ref{lemma:1-norm} and the fact that $\rho\leq 1$. We define
$$F(\xb):=-\nabla V(\theta(\xb))\cdot\ov D(\theta(\xb))+\zeta\ge 0\,,$$ 
and we rewrite inequality \eqref{estim-Lyap} as
\be\label{estim-Lyap2}\nabla V(\theta(\xb))\cdot D(\xb)\leq 
-F(\xb)+\zeta+\ov\epsilon\ .\ee
We now apply Proposition \ref{prop-approximate-Lyapunov1}. Concentration inequality \eqref{mu-conc-mf} together with Definition \eqref{Xdelta} and inequality \eqref{epsn} yield
\begin{align*}
&\ds  \mu \l(\mc X_{\delta-\zeta}\r)= \mu \l(\l\{\xb\in\mc X\,|\,F(\xb)<\delta\r\}\r) \\[10pt]
&\qquad\ge\ds1-\frac{K}{n\delta}-\frac{4||\nabla V||_{\infty}}{(n-1)\delta}-2|\mc A|^2||\nabla V||_{\infty}\frac{W_{\mc G}}{\delta}-\frac{\zeta}{\delta}\\[10pt]
&\qquad\ge\ds1-\frac{C}{n\delta}-2|\mc A|^2||\nabla V||_{\infty}\frac{W_{\mc G}}{\delta}-\frac{\zeta}{\delta}\,,
\end{align*}
with $C=K+{8}||\nabla V||_{\infty}$. 
\end{IEEEproof}\medskip


{
Notice that the right-hand side of the inequality \eqref{mu-conc-mf} is the sum of various terms. The first one depends only of the size on the interaction network and is negligible for large scale networks, the second one depends on the distance of the interaction network from a totally mixing one, while the last one depends on how far the function $V$ is from being a mean-field Lyapunov function for the PIN model.}


A particularly relevant special case of Theorem \ref{theo:general} is when \eqref{eq:Lyapunov-mf*} holds true with $\zeta=0$, i.e., when 
\be\label{eq:Lyapunov-mf222}\nabla V(\theta)\cdot  \ldrift(\theta)\leq 0\,,\qquad \forall \theta\in\mc P(\mc A)\,.\ee
In fact, a function $V$ satisfying \eqref{eq:Lyapunov-mf22} can be interpreted as a Lyapunov function for the deterministic mean-field ODE \eqref{ODE1}.
Furthermore, in the special case of complete interaction pattern, we have $\mc W_{\mc G}=0$ so that Theorem \ref{theo:general} has the following 
corollary.  

\begin{corollary}\label{cor:mfmodel} 
Consider a PIN model with complete interaction pattern and let $\mu$ in $\mc P(\mc X)$ be one of its stationary distributions. 
Assume that there exists class-$\mc C^2$  function $V:\mc P(\mc A)\to\R$ satisfying \eqref{eq:Lyapunov-mf22}. 
Then, there exists a constant $C$ only depending on $V$ such that 
\be\label{mu-conc-mf} \mu \l(\mc X_{\delta} \r)\ge1-\frac{C}{n\delta}\ee
for every  $\delta>0$.  
\end{corollary}

\subsection{Proofs of Theorems \ref{cor:ATM-general-2} and \ref{coro-approx-ER}}\label{sec:proofs}


%
%
Theorems \ref{cor:ATM-general-2} and \ref{coro-approx-ER} are a direct consequence of Theorem \ref{theo:general} applied to ATM interaction networks. Detailed proofs are below.

\begin{IEEEproof}[Proof of Theorem \ref{cor:ATM-general-2}]
It follows from Theorem \ref{theo:general} with $\zeta=0$ and Definition \ref{def:tfm} that
$$\mu_n\l(\mc X_{\delta}\r)\ge1-\frac{C}{n\delta}-|\mc A|^2||\nabla V||_{\infty}\frac{W_{\mc G_n}}{\delta}\stackrel{n\to+\infty}{\longrightarrow}1\,,$$
for every $\delta>0$. 
\end{IEEEproof}\medskip

\begin{IEEEproof} [Proof of Theorem \ref{coro-approx-ER} ]
Define
$$\mc X_{\delta}^{(0)}:=\{\xb\in\mc X:\,{ h(\theta(\xb))}<\delta\}\,,$$
and, for every $\alpha>0$,  
$$\mc X_{\delta}^{(\alpha)}:=\{\xb\in\mc X:\,\nabla V_\alpha(\theta(\xb))\cdot\ldrift_\alpha(\theta(\xb))>-\delta\}\,,$$
{As Theorem \ref{coro-approx-ER} assumes inequality \eqref{eq:Valpha}, we can apply Theorem \ref{theo:general} and obtain that
for every $\alpha>0$ and $\delta>\zeta(\alpha)$ the following estimation holds} 
$$
\ba{rcl}
\ds\mu_n^{(\alpha)}\left(\mc X_{\delta}^{(0)}\right)
&\ge&\ds\mu_n\l(\mc X_{\delta-\zeta(\alpha)}^{(\alpha)}\r)\\[7pt]
&\ge&\ds1-\frac{C}{n\delta}-|\mc A|^2||\nabla V||_{\infty}\frac{W_{\mc G_n}}{\delta}-\frac{\zeta(\alpha)}{\delta}\\[7pt]
&\!\!\!\!\!\stackrel{n\to+\infty}{\longrightarrow}\!\!\!\!\!&\ds1-\frac{\zeta(\alpha)}{\delta}\,,
\ea$$
{where the limit relation follows from the ATM assumption (see Definition \ref{def:tfm}).
This yields }
$$\liminf_{n\to+\infty}\ds\mu_n^{(\alpha)}\left(\mc X_{\delta}^{(0)}\right)\ge 1-\frac{\zeta(\alpha)}{\delta}\,.$$
It then follows from \eqref{limalpha} that 
$$\lim_{\alpha\to0}\liminf_{n\to+\infty}\ds\mu_n^{(\alpha)}\left(\mc X_{\delta}^{(0)}\right)\ge 1-\lim_{\alpha\to0}\frac{\zeta(\alpha)}{\delta}=1\,,$$
thus proving \eqref{mu-conc-mf2er}.\end{IEEEproof}

\section{Conclusion}\label{sec:conclusion}
In this paper, we have proposed a tool that provides estimates on the invariant distribution of Markov chains resulting from network systems in which agents change their state by a mutation or by an interaction with their neighbor agents. The key ingredient that enables such estimates is the existence of a Lyapunov type function for the drift associated with this system. This method is intrinsically robust and this permits its application to situations in which the model is only partially known and there exists a degree of uncertainty on it. We applied this method by first proving that, in some sense, the Erd\"os-R\'enyi random graphs are approximations of the complete graphs. Leveraging on this fact, we have been able to show 
%
%
that the asymptotic behavior of the SIS and SIRS epidemics models on Erd\"os-R\'enyi random graphs coincides with the asymptotic behavior of the SIS and SIRS models on the complete graphs. 

However, the range of applicability of this robust convergence result is much larger than this. Indeed, on the one hand, it can be applied to other families of Markov chains resulting from multi-agent systems, such as to Markov chains associated with evolutionary population games. On the other hand, it is possible to extend its application to different types of uncertainty, such as in cases in which the behavior of the agents is approximately described by a transition probability or even in the case in which a percentage of the agents have a completely unknown behavior. These kinds of extensions are the subject of our present investigations.

\bibliographystyle{unsrt}
\bibliography{bib}

\appendices

\section{Proof of Lemma \ref{lemma1}}\label{sec:proof:lemma1}

The fact that $\mu$ is stationary implies that 
\be\label{invariant}\E_{\mu}[V(\theta(X(t+1))-V(\theta(X(t))]=0\ee 
Since $V$ is $\mc C^2$, for every $\theta$ and $h$ in $\R^{\mc A}$, there exists $\beta$ in $[0,1]$ such that
$$V(\theta+h)=V(\theta)+\nabla V(\theta)\cdot h+\frac{1}{2}h^\top\Delta V(\theta+\beta h)h\ .$$
If we apply this expansion to both sides and use (\ref{invariant}) letting
$$\theta:=\theta(X(t))\,,\qquad h:=\theta(X(t+1))-\theta(X(t))\,,$$ 
we obtain
\be\label{taylor-equality}\E_\mu[\nabla V(\theta)h]=-\frac{1}{2}\E_\mu[h^\top\Delta V(\theta+\beta h)h]
\,.\ee
Notice now that, for every $\xb$ in $\mc X$, 
$$\ba{rcl}\ds\!\!\!\!\E_{\mu}\!\left[h|X(t)=\xb\right]
&\!\!\!\!=\!\!\!\!&\ds\E_{\mu}\!\left[\big(\theta(X(t+1))-\theta(X(t))\big)|X(t)=\xb\right]\\[7pt]
&\!\!\!\!=\!\!\!\!&\ds\frac1nD(\xb)\,,\ea$$
so that  the left-hand side of (\ref{taylor-equality}) is equal to
\be\label{estim1}\begin{array}{rcl}
\ds\E_\mu\left[\nabla V(\theta)\cdot h\right]
&=&\ds\E_\mu\left[\E[\nabla V (\theta)\cdot h|X(t)]\right]\\[7pt]
&=&\ds\E_\mu\left[\nabla V (\theta)\cdot\E[h|X(t)]\right]\\[7pt]
&=&\ds\frac1n\E_\mu\left[\nabla V (\theta)\cdot D(X(t))\right]\,.
\ea\ee
From \eqref{taylor-equality} and \eqref{estim1}, we obtain
\be\label{estim-bis}\ba{rcl}|\E_\mu\left[\nabla V (\theta)\cdot D(X(t))\right]|
&\!\!\leq\!\!&\ds\frac{n}{2}E_{\mu}\left[|h^\top\Delta V (\theta+\beta h)h|\r]\\[7pt]
&\!\!\leq\!\!&\ds\frac{n}{2} C_{ V }E_\mu\l[h^\top h\r]\,,
\ea
\ee
where $C_V=\max_{\theta} ||\Delta V(\theta)||_2$ and $||\cdot ||_2$ denotes the induced $2$-norm of a matrix.
Since $\|\theta({\bf y})-\theta(\xb)\|_2^2\le 2/n^2$ for every $\xb$ and $\bf y$ in $\mc X$ such that $\P(X(t+1)={\bf y}|X(t)=\xb)\not=0$, we have that $$E_\mu\l[h^\top h\r]\le\frac2{n^2}\,,$$
which, together with \eqref{estim-bis} implies the claim. $\qed$

\section{Proof of Proposition \ref{prop:Erdos-Renyi}}\label{proof:prop:Erdos-Renyi}

We start by recalling the following fundamental large-deviations result, known as Hoeffding-Azuma inequality \cite{Alon.Spencer:2008}. 

\begin{lemma}\label{lemma:Chernov-bound}
Let $Y$ be a random variable with binomial distribution with parameters $l$ and $p$. Then, for every $\alpha>0$
\be\label{HA-1}\P\left(Y\ge{lp}(1+\alpha)\right)\le \exp(-l\alpha^2/2)\,,\ee
and
\be\label{HA-2}\P\left(Y\le{lp}(1-\alpha)\right)\le \exp(-l\alpha^2/2)\,.\ee
\end{lemma}

\medskip

\begin{proposition}\label{prop:Erdos-Renyi-old}
Consider the Erd\"os-R\'enyi random graph $\mc G(n,p)$ with $n\ge2$ nodes and link probability $p$ in $]0,1]$. 
Then, 
\be\label{bound3}\P(W_{\mc G(n,p)}\ge\eta)\le 3^{n+2}\exp(-n^2p\eta^3/32)\,,\ee
for every { $0<\eta\le1$}. 
\end{proposition}
\begin{IEEEproof}
Let $Y=m/2$ be the number of undirected links in the Erd\"os-R\'enyi random graph $\mc G(n,p)$.  
Then $Y$ is a random variable  with binomial distribution with parameters $\binom n2$ and $p$. 
For $n\ge2$, we have $2(n-1)\ge n$, so that \eqref{HA-1} implies that
\be\label{M>}
\ba{rcl}
\ds\P\left(m\ge n(n-1)p(1+\alpha)\right)
&=&\ds\P\left(Y\ge \binom n2p(1+\alpha)\right)\\[7pt]
&\le&\ds\exp(- n(n-1)\alpha^2/4)\\[7pt]
&\le&\ds\exp(- n^2\alpha^2/8)\,,\ea\ee
while \eqref{HA-2} implies that 
\be\label{M<}\ba{rcl}
\ds\P\left(m\le n(n-1)p(1-\alpha)\right)
&=&\ds\P\left(Y\le  \binom n2p(1-\alpha)\right)\\[7pt]
&\le&\ds\exp(- n(n-1)\alpha^2/4)\\[7pt]
&\le&\ds\exp(- n^2\alpha^2/8)\,.\ea\ee
On the other hand, let 
$$\mc W=\left\{(\mc S,\mc U):\,\mc S\subseteq\mc V,\mc U\subseteq\mc V,\mc S\cap\mc U=\emptyset\right\}\,.$$
Then, for $(\mc S,\mc U)$ in $\mc W$, 
the number $|\mc E_{\mc S\mc U}|$ of directed links from nodes in $\mc S$ to nodes in $\mc U$ in $\mc G(n,p)$, that is equal to
the number of undirected links connecting nodes in $\mc S$ with nodes in $\mc U$, is a binomial random variable of parameters $|\mc S||\mc U|$ and $p$, so that, for every $\alpha\ge0$,  
\be\label{Mab>}\P(|\mc E_{\mc S\mc U}|\ge |\mc S||\mc U|p(1+\alpha))\le\exp(- |\mc S||\mc U|{\alpha^2}/{2})\,.\ee

Fixed any $\eta$ such that $0<\eta\le 1$, define 
$$\mc W_{\eta}:=\left\{(\mc S,\mc U)\in\mc W:\,\frac{|\mc S||\mc U|}{n(n-1)}\le \frac{p\eta}2\right\},\ \  \ov{\mc W}_{\eta}=\mc W\setminus\mc W_{\eta}\,.$$
Notice that, for every $(\mc S,\mc U)$ in $\mc W_{\eta}$,  
$$|\mc E_{\mc S\mc U}|\le|\mc S||\mc U|\le n(n-1) p\eta/2\,,$$
so that, using also the fact that $m\le n(n-1)$, we get
$$\left|\frac{|\mc E_{\mc S\mc U}|}{m}-\frac{|\mc S||\mc U|}{n(n-1)}\right|\le
\frac{|\mc S||\mc U|}{m}\le
\frac{n(n-1) p\eta}{2m}=\binom{n}{2}\frac{p\eta}m\,.$$
Therefore, 
\be\label{bound1}\ba{rcl}
\ds\P\left(\max\limits_{(\mc S,\mc U)\in\mc W_{\eta}}\left|\frac{|\mc E_{\mc S\mc U}|}{m}-\frac{|\mc S||\mc U|}{n(n-1)}\right|\ge\eta\right)
&\le&\P\left(m\le \binom n2p\right)\\[8pt]
&\le&\exp(-n^2/32)\,,
\ea\ee
where the second inequality follows from \eqref{M<} with $\alpha=1/2$. 
On the other hand, for $(\mc S,\mc U)$ in $\ov{\mc W}_{\eta}$, define 
\begin{align*}
p_{\mc S\mc U}^+&:=\P\left(\frac{|\mc E_{\mc S\mc U}|}{m}-\frac{|\mc S||\mc U|}{n(n-1)}\ge\eta\right)\\
p_{\mc S\mc U}^-&:=\P\left(\frac{|\mc E_{\mc S\mc U}|}{m}-\frac{|\mc S||\mc U|}{n(n-1)}\le-\eta\right)\,,
\end{align*}
and observe that 
{
\be\label{p+}\ba{rcl}p^+_{\mc S\mc U}
&=&\ds\P\left(\frac{|\mc E_{\mc S\mc U}|n(n-1)}{m|\mc S||\mc U|}\ge1+\eta \frac{n(n-1)}{|\mc S||\mc U|}\right)\\[10pt]
&\le&\ds\P\left(\frac{|\mc E_{\mc S\mc U}|n(n-1)}{m|\mc S||\mc U|}\ge1+2\eta\right)\\[10pt]
&\le&\ds\P\left(\frac{|\mc E_{\mc S\mc U}|n(n-1)}{m|\mc S||\mc U|}\ge\frac{1+\eta/2}{1-\eta/2}\right)\\[10pt]
&\le&\ds\P\left(\Big\{|\mc E_{\mc S\mc U}|\ge p|\mc S||\mc U|(1+\eta/2)\Big\}\right.\\[5pt]
&&\ds\ \qquad\qquad \cup\left.\Big\{m\le n(n-1)p(1-\eta/2)\Big\}\right)\\[10pt]
&\le&\ds\P\left(|\mc E_{\mc S\mc U}|\ge p|\mc S||\mc U|(1+\eta/2)\right)\\[5pt]
&&\qquad\quad\ \ds+\P\left(m\le n(n-1)p(1-\eta/2)\right)\\[10pt]
&\le&\exp(-|\mc S||\mc U|\eta^2/8)+\exp(-n^2\eta^2/32)\\[10pt]
&\le&\exp(-n^2p\eta^3/32)+\exp(-n^2\eta^2/32)\\[10pt]
&\le&2\exp(-n^2p\eta^3/32)\,,\ea\ee
where the first inequality follows from the fact that 
\be\label{theta<1/14}|\mc S||\mc U|\le|\mc S|(n-|\mc S|)\le n^2/4\,,\ee the second one since 
$$\frac{1+\eta/2}{1-\eta/2}=1+\frac{\eta}{1-\eta/2}\le1+2\eta\,,\qquad \forall \eta\ : 0\le\eta\le 1\,,$$
the forth one from a union bound argument, the fifth one by applying \eqref{Mab>} and \eqref{M<} with $\alpha=\eta/2$, respectively, the sixth one since 
$$|\mc S||\mc U|> n^2p\eta/4\,,$$ 
for every $(\mc S,\mc U)\notin\mc W_{\eta}$, and the last one since $p\eta\le1$. 
}

Similarly, we get 
{
\be\label{p-}\ba{rcl}p^-_{\mc S\mc U}
&=&\ds\P\left(\frac{|\mc E_{\mc S\mc U}|n(n-1)}{m|\mc S||\mc U|}\le 1-\eta \frac{n(n-1)}{|\mc S||\mc U|}\right)\\[10pt]
&\le&\ds\P\left(\frac{|\mc E_{\mc S\mc U}|n(n-1)}{m|\mc S||\mc U|}\le1-2\eta\right)\\[10pt]
&\le&\ds\P\left(\frac{|\mc E_{\mc S\mc U}|n(n-1)}{m|\mc S||\mc U|}\ge\frac{1-\eta}{1+\eta}\right)\\[10pt]
&\le&\ds\P\left(\Big\{|\mc E_{\mc S\mc U}|\le p|\mc S||\mc U|(1-\eta)\Big\}\right.\\[5pt]
&&\ds\ \qquad\qquad \cup\left.\Big\{m\ge n(n-1)p(1+\eta)\Big\}\right)\\[10pt]
&\le&\ds\P\left(|\mc E_{\mc S\mc U}|\le p|\mc S||\mc U|(1-\eta)\right)\\[5pt]
&&\qquad\quad\ \ds+\P\left(m\ge n(n-1)p(1+\eta)\right)\\[10pt]
&\le&\exp(-|\mc S||\mc U|\eta^2/2)+\exp(-n^2\eta^2/8)\\[10pt]
&\le&\exp(-n^2p\eta^3/8)+\exp(-n^2\eta^2/8)\\[10pt]
&\le&2\exp(-n^2p\eta^3/8)\,,\ea\ee
where the first inequality follows from the fact that 
\be\label{theta<1/14}|\mc S||\mc U|\le|\mc S|(n-|\mc S|)\le n^2/4\,,\ee the second one since 
$$\frac{1-\eta}{1+\eta}=1-\frac{2\eta}{1+\eta}\ge1-2\eta\,,\qquad \forall \eta\ : 0\le\eta\le 1\,,$$
the forth one from a union bound argument , the fifth one by applying \eqref{Mab>} and \eqref{M<} with $\alpha=\eta$, respectively, the sixth one since 
$$|\mc S||\mc U|> n^2p\eta/4\,,$$ 
for every $(\mc S,\mc U)\notin\mc W_{\eta}$, and the last one since $p\eta\le1$. 
}



For every $(\mc S,\mc U)$ in $\ov{\mc W}_{\eta}$, let 
$$p_{\mc S\mc U}=\P\left(\left|\frac{|\mc E_{\mc S\mc U}|}{2m}-\frac{|\mc S||\mc U|}{n(n-1)}\right|\ge\eta\right)\,.$$
{A union bound argument together with inequalities \eqref{p+} and \eqref{p-}} imply that 
$$p_{\mc S\mc U}\le p_{\mc S\mc U}^++p_{\mc S\mc U}^-\le4\exp(-n^2p\eta^3/32)\,.$$
{By now applying one more time the union bound, the above estimation}, and noting that $|\ov{\mc W}_{\eta}|\le|\mc W|=3^n$, we get 
\be\label{bound2}
\ba{rcl}
&&\ds\P\left(\max_{(\mc S,\mc U)\in\mc W_{\eta}}\left|\frac{|\mc E_{\mc S\mc U}|}{2m}-\frac{|\mc S||\mc U|}{n(n-1)}\right|\ge\eta\right)\\[10pt]
&&\qquad\le\ds\sum_{(\mc S,\mc U)\in\mc W_{\eta}}p_{\mc S\mc U}
\le\ds 4|\mc W_{\eta}|\exp(-n^2p\eta^3/32)\\[10pt]
&&\qquad\le4\cdot3^n\exp(-n^2p\eta^3/32)\,.\ea\ee 

Then, estimations \eqref{bound1}  and  \eqref{bound2} imply that 
\begin{align*}
\P\left(W_{\mc G(n,p)}\ge\eta\right)
&=\P\left(\max_{(\mc S,\mc U)\in\mc W}\left|\frac{|\mc E_{\mc S\mc U}|}{2m}-\frac{|\mc S||\mc U|}{n(n-1)}\right|\ge\eta\right)\\[8pt] 
&\le\exp(-n^2/32)+4\cdot3^n\exp(-n^2p\eta^3/32)\\[8pt]
&\le (4\cdot3^n+1)\exp(-n^2p\eta^3/32)\\[8pt]
&\le 3^{n+2}\exp(-n^2p\eta^3/32)\,, 
\end{align*} 
thus proving the claim. 
\end{IEEEproof}\medskip

We can now easily prove Proposition \ref{prop:Erdos-Renyi} from Proposition  \ref{prop:Erdos-Renyi-old}.

\begin{IEEEproof}[Proof of Proposition \ref{prop:Erdos-Renyi}]
If $np\to+\infty$, then we have that,  
for every $\eta>0$, the expresssion $3^{n+2}\exp(-n^2p\eta^3/32)$ vanishes faster than exponentially as $n$ grows large. 
Hence, by \eqref{bound3}, 
$$\sum_{n\ge2}\P(W_{\mc G(n,p)}\ge\eta)\le\sum_{n\ge2} 3^{n+2}\exp(-n^2p\eta^3/32)<+\infty\,,$$
and the Borel-Cantelli lemma implies that 
$$\limsup_{n\to+\infty}W_{\mc G(n,p)}\le\eta\,$$
with probability $1$.
As this holds true for every $\eta>0$, we conclude that $W_{\mc G(n,p)}\to0$ with probability $1$. \end{IEEEproof}

\newpage
\begin{IEEEbiography}
[{\includegraphics[width=1in,height=1.25in,clip,keepaspectratio]{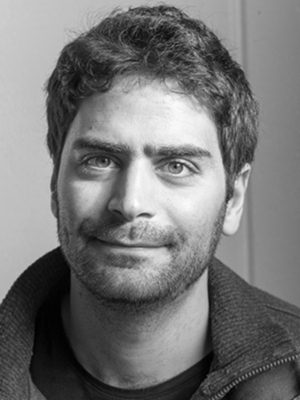}}]
{Giacomo Como}(M'12) is  a  Professor at  the Department  of  Mathematical  Sciences,  Politecnico di  Torino,  Italy. He is also a Senior Lecturer  at  the  Automatic  Control  Department, Lund  University,  Sweden.  He  received the B.Sc., M.S., and Ph.D.~degrees in Applied Mathematics  from  Politecnico  di  Torino,  in  2002,  2004, and 2008, respectively. He was a Visiting Assistant in  Research  at  Yale  University  in  2006--2007  and  a Postdoctoral  Associate  at  the  Laboratory  for  Information  and  Decision  Systems,  Massachusetts  Institute of Technology, from 2008 to 2011. Prof.~Como currently serves  as Senior Editor of the  \textit{IEEE Transactions on Control of Network Systems} and as Associate  Editor  of \textit{Automatica}. He is the chair  of the  {IEEE-CSS  Technical  Committee  on  Networks  and  Communications}. He served as Associate Editor for the  \textit{IEEE Transactions on Network Science and Engineering} (2015-2021) and of the \textit{IEEE Transactions on Control of Network Systems} (2016-2022).  He was  the  IPC  chair  of  the  IFAC  Workshop  NecSys'15  and  a  semiplenary speaker  at  the  International  Symposium  MTNS'16.  He  is  recipient  of  the 2015  George S.~Axelby  Outstanding Paper Award.  His  research interests  are in  dynamics,  information,  and  control  in  network  systems  with  applications to  cyber-physical  systems,  infrastructure  networks,  and  social  and  economic networks.\end{IEEEbiography}

\begin{IEEEbiography}
[{\includegraphics[width=1in,height=1.25in,clip,keepaspectratio]{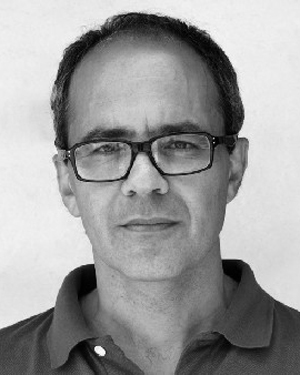}}]
{Fabio Fagnani}
received the Laurea degree in Mathematics from the University of Pisa and the Scuola Normale Superiore, Pisa, Italy, in 1986. He received the PhD degree in Mathematics from the University of Groningen,  Groningen,  The  Netherlands,  in 1991. From 1991 to 1998, he was an Assistant Professor of Mathematical Analysis at the Scuola Normale Superiore. In 1997, he was a Visiting Professor at the Massachusetts Institute of Technology, Cambridge MA. Since 1998, he has been with the Politecnico of Torino, where he is currently (since 2002) a Full Professor of Mathematical Analysis. From 2006 to 2012, he has acted as Coordinator of the PhD program in Mathematics for Engineering Sciences at Politecnico di Torino. From June 2012 to September 2019, he served as the Head of the Department of Mathematical Sciences, Politecnico di Torino. His current research topics are on cooperative algorithms and dynamical systems over graphs, inferential distributed algorithms, and opinion dynamics. He is an Associate Editor of the \textit{IEEE Transactions on Automatic Control} and served in the same role for the \textit{IEEE Transactions on Network Science and Engineering} and \textit{IEEE Transactions on Control of Network Systems}.
\end{IEEEbiography}

\begin{IEEEbiography}
[{\includegraphics[width=1in]{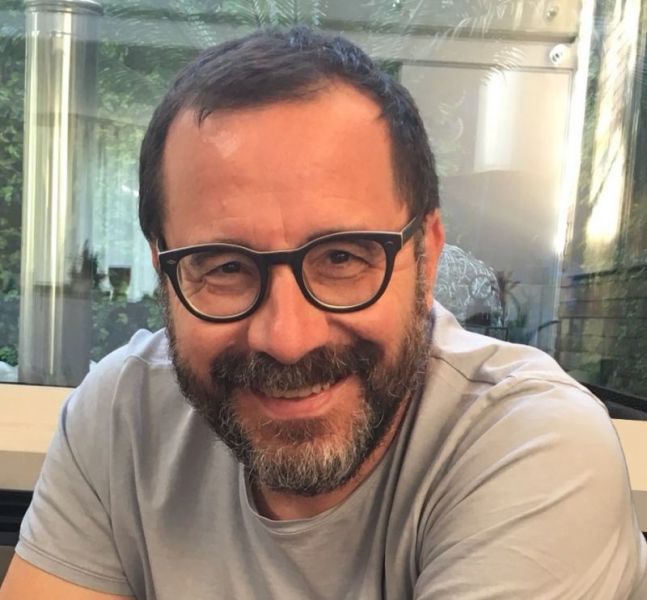}}]
{Sandro Zampieri} received the Laurea degree in Electrical Engineering and the Ph.D.~degree in System Engineering from the University of Padova, Italy, in 1988 and 1993, respectively. Since 2002 he has been Full Professor of Automatic Control at the Department of Information Engineering of the University of Padova. He has been the head of the Department of Information Engineering from 2014 until 2018. In 1991--1992, 1993 and 1996 he was Visiting Scholar at the Laboratory for Information and Decision Systems,
Massachusetts Institute of Technology. He has held visiting positions also at the Department of Mathematics of the University of Groningen and at the Department of Mechanical Engineering of the University of California at Santa Barbara.
Prof. Zampieri was the general chair of the 1st IFAC Workshop on Estimation and Control of Networked Systems 2009, program chair of the 3rd IFAC Workshop on Estimation and Control of Networked Systems 2012 and publication chair of the IFAC World Congress 2011. He was the chair of the IFAC technical committee ”Networked systems” on 2005--2008. He was one of the recipients of the 2016 IEEE Transactions on Control of Network Systems Best Paper Award. He has been an IEEE Fellow since 2022. His research interests include networked control, control of complex systems and distributed control and estimation with applications to the smart grids.\end{IEEEbiography}

\end{document}